\documentclass[%
 reprint,
 amsmath,amssymb,
 aps,
]{revtex4-2}

\usepackage{graphicx}% Include figure files
\usepackage{dcolumn}% Align table columns on decimal point
\usepackage{bm}% bold math
\usepackage{multirow}
\usepackage{lipsum}
\begin{document}

\preprint{APS/123-QED}

\title{How does Goldene Stack?}

\author{Marcelo Lopes Pereira, Jr}
\affiliation{University of Bras\'{i}lia, College of Technology, Department of Electrical Engineering, Bras\'{i}lia, Federal District, Brazil.}
\author{Emanuel J. A. dos Santos}
 \affiliation{Institute of Physics, University of Bras\'ilia, 70910-900, Bras\'ilia, Brazil}
\affiliation{Computational Materials Laboratory, LCCMat, Institute of Physics, University of Bras\'ilia, 70910-900, Bras\'ilia, Brazil}
\author{Luiz Antonio Ribeiro, Jr}
 \affiliation{Institute of Physics, University of Bras\'ilia, 70910-900, Bras\'ilia, Brazil}
\affiliation{Computational Materials Laboratory, LCCMat, Institute of Physics, University of Bras\'ilia, 70910-900, Bras\'ilia, Brazil}
\author{Douglas Soares Galvão}
\affiliation{Department of Applied Physics and Center for Computational Engineering and Sciences, State University of Campinas, Campinas, São Paulo, Brazil.}

\date{\today}

\begin{abstract}
The recent synthesis of Goldene, a 2D atomic monolayer of gold, has opened new avenues in exploring novel materials. However, the question of when multilayer Goldene transitions into bulk gold remains unresolved. This study used density functional theory calculations to address this fundamental question. Our findings reveal that multilayer Goldene retains an AA-like stacking configuration of up to six layers, with no observation of Bernal-like stacking as seen in graphene. Goldene spontaneously transitions to a bulk-like gold structure at seven layers, adopting a rhombohedral (ABC-like) stacking characteristic of bulk face-centered cubic (FCC) gold. The atomic arrangement converges entirely to the bulk gold lattice for more than ten layers. Quantum confinement significantly impacts the electronic properties, with monolayer and bulk Goldene exhibiting a single Dirac cone at the X-point of the Brillouin zone. In contrast, multilayer Goldene shows two Dirac cones at the X- and Y-points. Additionally, monolayer Goldene exhibits anisotropic optical absorption, which is absent in bulk gold. This study provides a deeper understanding of multilayer Goldene's structural and electronic properties and stacked 2D materials in general.
\end{abstract}

\maketitle

%\tableofcontents

\section{Introduction} 

The active research in the growing class of two-dimensional (2D) materials has revolutionized our understanding of condensed matter physics, providing new platforms for exploring exotic electronic, optical, and mechanical properties in flat electronics \cite{zeng2018exploring,xu2013graphene}. Among these materials, graphene, a single layer of carbon atoms arranged in a honeycomb lattice, has started a revolution due to its remarkable properties, including high carrier mobility \cite{geim2009graphene,geim2007rise}. Graphene's ability to stack into multilayers, transitioning from monolayer graphene to bulk graphite, has been extensively studied \cite{yang2019stacking,wang2009vibrational,liu2013temperature,wang2019ultrastiff,wang2016stacking}, revealing distinct stacking behaviors such as Bernal (AB-like) and rhombohedral (ABC-like) configurations \cite{nery2020long,xu2012pathway,sugawara2018selective,zhou2021half,zhou2021superconductivity}. These stacking arrangements critically influence the electronic structure, leading to tunable properties that have implications for fundamental research and applications \cite{latychevskaia2019stacking}.

Raman spectroscopy has been a vital tool in probing the transition from multilayered graphene to graphite \cite{ferrari2013raman}. One of the characteristic features of graphene's vibrational properties is the G-band, which corresponds to the in-plane vibrations of sp$^{2}$ carbon atoms. Studies have shown that as the number of stacked graphene layers increases, the G-band gradually converges to a value characteristic of graphite \cite{liu2013temperature}. Precisely, for multilayered graphene, the G-band shifts to 1581.6 $\pm$ 1.1 cm$^{-1}$ when up to 10 graphene layers are stacked, a value consistent with bulk graphite \cite{wang2009vibrational}. This convergence signifies the transition of graphene's electronic and vibrational properties towards those of graphite. Notably, the electronic transport properties of few-layer graphene follow a temperature dependence that further corroborates this transition to graphite-like behavior as the number of layers increases \cite{liu2013temperature}. These findings emphasize how multilayer stacking affects the physical properties of 2D materials, encouraging studies for elucidating the stacking transitions in other nanomaterials.

In recent years, attention has increasingly shifted toward a broader class of 2D nanomaterials, expanding the possibilities far beyond graphene’s well-established properties \cite{hou2022synthesis,desyatkin2022scalable,hu2022synthesis,fan2021biphenylene,meirzadeh2023few,toh2020synthesis}. One fascinating advancement was the discovery of Goldene, a monolayer of gold atoms that marks a breakthrough in the fabrication of 2D metals \cite{kashiwaya2024synthesis}. While previous efforts to create 2D gold were limited to multi-atom-thick films or confined monolayers, Goldene represents a real single-atom-thick structure. It was synthesized through a wet-chemical etching process that selectively removes Ti$_3$C$_2$ layers from a nanolaminate precursor, Ti$_3$AuC$_2$, a compound derived by substituting gold for silicon in the well-known MAX phase material Ti$_3$SiC$_2$. This process results in a Goldene layer that exhibits a 9\% lattice contraction compared to bulk gold, as confirmed by electron microscopy \cite{kashiwaya2024synthesis}.

Overcoming challenges in synthesizing this novel material --- such as curling and agglomeration --- was made possible through the strategic use of surfactants, which play a crucial role in stabilizing the exfoliated layers \cite{kashiwaya2024synthesis}. Further X-ray photoelectron spectroscopy investigations have revealed a notable shift in the Au 4f binding energy, increasing by 0.88 eV relative to bulk gold, highlighting Goldene’s unique electronic properties \cite{kashiwaya2024synthesis,zhao2024electrical}. Importantly, molecular dynamics simulations show that Goldene is inherently stable at the atomic level, sparking interest in scaling up its production for further exploration \cite{kashiwaya2024synthesis,mortazavi2024goldene}. Moreover, other theoretical studies have also pointed for the stability of quasi-2D gold clusters \cite{yoon2007size,koskinen2006density,koskinen2007liquid,stiehler2013electron,koskinen2015plenty}. As a new addition to the growing family of 2D metals, Goldene presents intriguing opportunities for exploring quantum confinement effects and electronic behavior, particularly concerning its stacking patterns, which are central to our investigation. In contrast to graphene, whose multilayer forms naturally adopt Bernal or rhombohedral stacking as they approach graphite, Goldene’s stacking behavior remains less understood, especially as it transitions towards the face-centered cubic (FCC) structure typical of bulk gold.

This work addresses a previously unanswered question: When does multilayer Goldene transition into bulk gold? Understanding this transition is essential for Goldene and a broader range of 2D materials where stacking plays a pivotal role in determining properties. By using density functional theory (DFT) calculations, we systematically investigated the structural evolution of Goldene as a function of the number of layers, revealing distinct stacking configurations and their electronic consequences. Goldene maintains an AA-like stacking of up to six layers, unlike graphene, without exhibiting the Bernal stacking characteristic of multilayer graphene. Interestingly, a transition to a bulk-like FCC structure occurs at seven layers, where rhombohedral stacking emerges naturally, mirroring the behavior of bulk gold. Beyond this, the lattice arrangement fully converges to that of bulk gold for more than ten layers. The electronic properties of multilayer Goldene also exhibit unique characteristics. While monolayer Goldene and bulk gold share a Dirac cone at the X-point of the Brillouin zone, multilayered Goldene shows two Dirac cones at both the X- and Y-points. Furthermore, monolayer Goldene displays anisotropic optical absorption, which vanishes in the bulk. These findings highlight the impact of quantum confinement, structural arrangement, and anisotropic electronic pathways in stacked Goldene layers, adding to the growing understanding of 2D materials and their multilayered analogs.

\section{Methodology}

\begin{figure*}[!htb]
    \centering
    \includegraphics[width=0.8\linewidth]{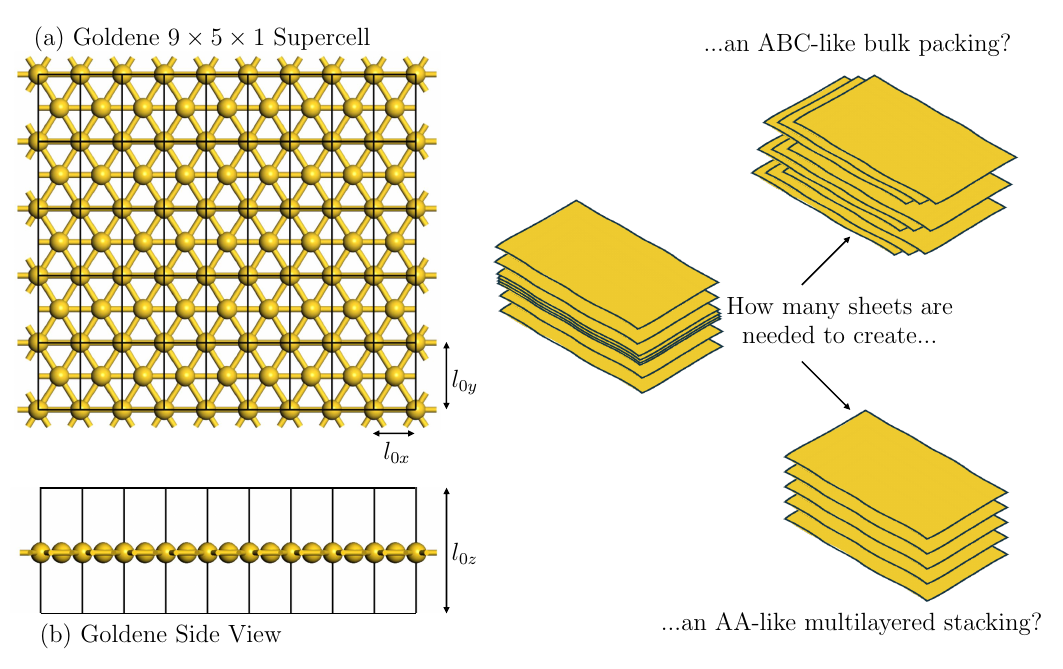}
    \caption{Schematic representation of Goldene: Front view (a) of a $9\times5\times1$ supercell, highlighting the unit cell replications and the magnitudes of the lattice vectors $ l_{0x} $ and $ l_{0y} $. Side view (b) of the Goldene monolayer, indicating the lattice parameter in the out-of-plane direction $ l_{0z} $. On the right, the main objective of this work: to determine how the stacking of Goldene monolayers depends on the number of layers.}
    \label{fig:system}
\end{figure*}

Based on the DFT formalism \cite{Kohn1965,Sanchez1997}, we studied the structural, electronic, and optical characteristics of Goldene multilayers using first-principles calculations. Figure \ref{fig:system}(a) shows a $9\times5\times1$ supercell of Goldene, with unit cells highlighted in smaller black rectangles. Each unit cell of Goldene consists of two atoms in a triangular lattice, with the lattice vectors' moduli indicated as $l_{0x}$ and $l_{0y}$. Figure \ref{fig:system}(b) provides a side view of Goldene, where the box size along the $z$ direction is marked as $l_{0z}$. In our investigation, the monolayer Goldene was initially optimized, and subsequently, we stacked two up to ten layers of Goldene, each separated by 5 \r{A}. The box size along the $z$ direction was fixed at 100 \r{A} to prevent interaction between the Goldene monolayers and their respective out-of-plane mirror images. Thus, the initial vacuum space varied from 60 \r{A} (for ten layers) to 100 \r{A} (for a monolayer). Periodic boundary conditions were applied in all directions. For comparison, we also investigated the bulk Goldene, setting $l_{0z} = 10$ \r{A}, with two layers of the Au-based system, without any restrictions on the lattice vectors or angles. The right part of Figure \ref{fig:system} exemplifies one of the main objectives of this work, which is to identify how the Goldene monolayers organize themselves, whether via AA-type stacking or in its bulk form with ABC interlayer arrangement.

The geometry optimization of Goldene multilayers, as well as the structural stability and electronic and optical properties, were obtained using first-principles calculations with the Spanish Initiative for Electronic Simulations with Thousands of Atoms (SIESTA) code \cite{Soler2002,Garcia2020}. In the Goldene multilayered systems, van der Waals (vdW) corrections were used to describe the exchange-correlation term \cite{dion2004van,thonhauser2007van}. The calculations were carried out within the generalized gradient approximation (GGA) framework \cite{ernzerhof1999assessment} using the Perdew-Burke-Ernzerhof (PBE) functional \cite{Perdew1996}.

Electron-ion interactions were described using Troullier-Martins norm-conserving pseudopotentials with typical valence electron configurations for Au atoms, in Kleinman-Bylander form \cite{Troullier1991,Kleinman1982}. An energy cutoff of 300 Ry and a double-$\zeta$ (DZP) basis set composed of numerical atomic orbitals with finite range were used in all calculations. We adopted the Monkhorst-Pack $5 \times 5 \times n$ grid \cite{Monkhorst1976} to integrate the Brillouin zone during optimization, while for obtaining electronic properties, a Monkhorst-Pack $30 \times 30 \times 3n$ grid was used, where $n$ is the number of Goldene layers. During optimization, the lattice vectors and atomic positions were fully relaxed until the maximum force on each atom was less than 0.001 eV/\text{\AA} and the total energy difference was less than $10^{-5}$ eV. 

To understand the stability of Goldene multilayer systems, the cohesive energy as a function of $n$ layers ($E_\text{coh}(n)$) was obtained from the expression: 
\begin{equation}
E_\text{coh}(n) = \frac{E_\text{$n$-layers}-N_\text{Au} E_\text{Au}}{N_\text{Au}}, 
\end{equation}
where $E_\text{$n$-layers}$, $N_\text{Au}$, and $E_\text{Au}$ represent the total energy of the complex with $n$ layers, the number of Au atoms in the system, and the power of a single isolated gold atom, respectively.

Phonon calculations were also carried out to investigate the Goldene monolayer's mechanical strength and confirm the system's stability. This approach allows us to identify vibrational modes with imaginary (negative) frequencies that indicate dynamic instability in the monolayer. These calculations used a $9 \times 5 \times 1$ supercell interpolated into the Brillouin zone with a mesh cutoff of 700 Ry. Convergence parameters were set to $10^{-5}$ for energy and 0.001 eV/\AA\ for force. The acoustic sum rule was also applied to vibrational frequencies at the $\Gamma$ point.

Finally, we also evaluated the influence of the bulk (ABC-like) and AA-like stacking arrangements on the optical properties of Goldene. Thus, we applied a standard external electric field of 1.0 V/\r{A} along the $x$-, $y$-, and $z$-directions separately to perform optical calculations. Using the Kramers-Kronig relation \cite{kramers1927,kronig1926} and Fermi's golden rule \cite{tignon1995}, we derived the real $(\epsilon_1)$ and imaginary $(\epsilon_2)$ parts of the dielectric constant. The real part is given by: 

\begin{equation}
\epsilon_1(\omega)=1+\frac{1}{\pi}P\int_{0}^{\infty}d\omega'\frac{\omega'\epsilon_2(\omega')}{\omega'^{2}-\omega^{2}},
\end{equation} 

where $P$ denotes the principal value of the integral over $\omega'$, the imaginary part, which considers interband optical transitions between the valence band (VB) and the conduction band (CB), is expressed as:

\begin{equation}
\epsilon_2(\omega)=\frac{4\pi^2}{V_\Omega\omega^2} \sum_{\substack{i\in \text{VB}\ j\in \text{CB}}} \sum_{k}W_k \left | \rho_{ij} \right |^2 \delta (\epsilon_{kj}-\epsilon_{ki}-\hbar\omega).
\end{equation}

In this equation, $\omega$ represents the photon frequency, $\left | \rho_{ij} \right |$ is the dipole transition matrix element, and $W_k$ is the weight of the respective $k$-point in reciprocal space. Additionally, $V_\Omega$ is the system volume, calculated as $V_\Omega = l_{0x} \cdot l_{0y} \cdot h_0$, where $h_0$ is the thickness of the Goldene monolayer, in the bulk case, the $l_{0z}$ was used instead of $h_0$. Using $\epsilon_1$ and $\epsilon_2$, the absorption coefficient ($\alpha$) was calculated from the expression:

\begin{equation}
\alpha(\omega)=\sqrt{2}\omega\left [ (\epsilon_1^{2}(\omega)+\epsilon_2^{2}(\omega))^{1/2}-\epsilon_1(\omega)\right ]^{1/2}.
\end{equation}

\section{Results}

\begin{figure}[t]
    \centering
    \includegraphics[width=8cm]{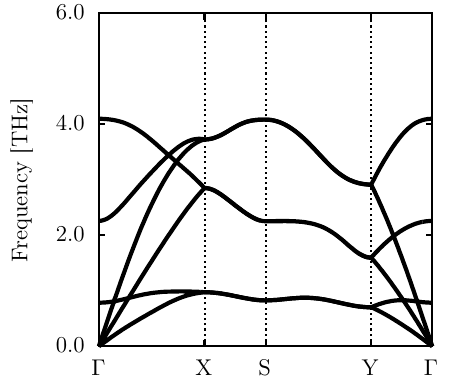}
    \caption{Phonon dispersion curves for the Goldene supercell $9\times5\times1$.}
    \label{fig:phonons}
\end{figure}

We begin by discussing the stability of Goldene's monolayer and the accuracy of our parameters for the calculations performed here. The phonon dispersion curves for monolayer Goldene, as shown in Figure \ref{fig:phonons}, confirm its dynamical stability in the fully flat configuration. As with many 2D materials, including graphene \cite{falkovsky2007phonon,koukaras2015phonon}, goldene exhibits three acoustic modes from the $\Gamma$ point, which are the longitudinal (LA), transverse (TA), and out-of-plane (ZA) acoustic modes, corresponding to its two-atom primitive cell. Three optical phonon modes (LO, TO, and ZO branches) were also observed due to the same atomic configuration. Interestingly, in contrast to lighter 2D materials like graphene, goldene's phonon spectrum spans a considerably lower frequency range. This difference highlights the substantial impact of atomic mass on phonon dynamics, as the heavier gold atoms contribute to a suppressed phonon group velocity. As a result, lattice thermal conductivity in goldene may be much lower compared to graphene, aligning with expectations for a 2D material composed of heavier elements. It is worthwhile to stress that similar results were obtained using different DFT approaches \cite{zhao2024electrical,mortazavi2024goldene} and also an MD framework based on machine learning interatomic potentials \cite{mortazavi2024goldene}, confirming the accuracy of the DFT approach employed in this study.

The phonon spectrum shows no imaginary frequencies, reinforcing that Goldene maintains a stable 2D lattice structure. This stability is crucial for applications in nanodevices, as it ensures the material's structural integrity under various conditions. Moreover, the relatively low Debye frequency of the acoustic phonons points to the inherently slow propagation of lattice vibrations, further distinguishing Goldene from lighter 2D materials.

Comparing the phonon dispersion of monolayer Goldene to that of bulk gold (in its FCC configuration) \cite{thakur1986lattice,dal2013ab} reveals notable similarities, especially in the frequency range. The phonon modes in bulk gold extend about 5 THz \cite{thakur1986lattice,dal2013ab}, very close to the Goldene one (see Figure \ref{fig:phonons}). The three-dimensional lattice and related vibrational modes do not influence the magnitudes of the phonon modes. However, the dispersion trend differs. These similarities and contrasts between the monolayer and bulk structures underscore how quantum confinement in Goldene impacts phononic behavior. In particular, the confinement in the 2D form reduces the frequency range since the highest phonon mode in Goldene is about 4 THz (see Figure \ref{fig:phonons}). They also lead to altered heat transport properties, a critical factor in understanding the thermal behavior of nanoscale materials.

\begin{figure}[t]
    \centering
    \includegraphics[width=8cm]{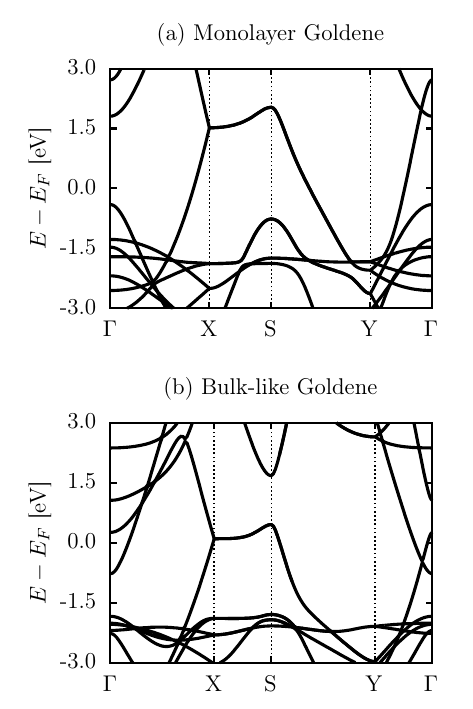}
    \caption{The electronic band structure of (a) the Goldene monolayer and (b) the bulk-like Goldene model.}
    \label{fig:bands01}
\end{figure}

We now discuss Goldene's electronic band structure. Figures \ref{fig:bands01}(a) and \ref{fig:bands01}(b) show the electronic band structure for the monolayer Goldene and the bulk-like Goldene, respectively. This bulk-like phase is modeled as a Goldene double layer AA-like stacking with a $l_{0z}=4.91$ \r{A}. The electronic band structure of monolayer and bulk-like Goldene, as presented in Figure \ref{fig:bands01}, offers essential insights into the material's conductive behavior. A distinct metallic nature is observed for the monolayer Goldene (see Figure \ref{fig:bands01}(a)), characterized by a single band crossing the Fermi level along high-symmetry paths in the Brillouin zone. This trend indicates that Goldene maintains a strong conductive characteristic even in its monolayer form. Moreover, the dispersion of the band near the Fermi level is quite steep, reflecting a high Fermi velocity. This feature suggests that Goldene has excellent electronic transport properties, which may rival or exceed those of other well-known 2D conductive materials, such as graphene. 

In the case of the bulk-like Goldene (see Figure \ref{fig:bands01}(b)), one can observe a similarly metallic band structure. However, a key difference emerges when comparing the position of the Dirac cone between the monolayer and bulk-like cases. For the monolayer, the Dirac cone appears at approximately 1.5 eV. In the bulk-like configuration, the Dirac point shifts to the Fermi level. This shift can be attributed to the interactions between layers in the bulk-like case, which lead to modifications in the band structure due to weak interlayer vdW coupling. 

\begin{figure*}[t]
    \centering
    \includegraphics[width=16cm]{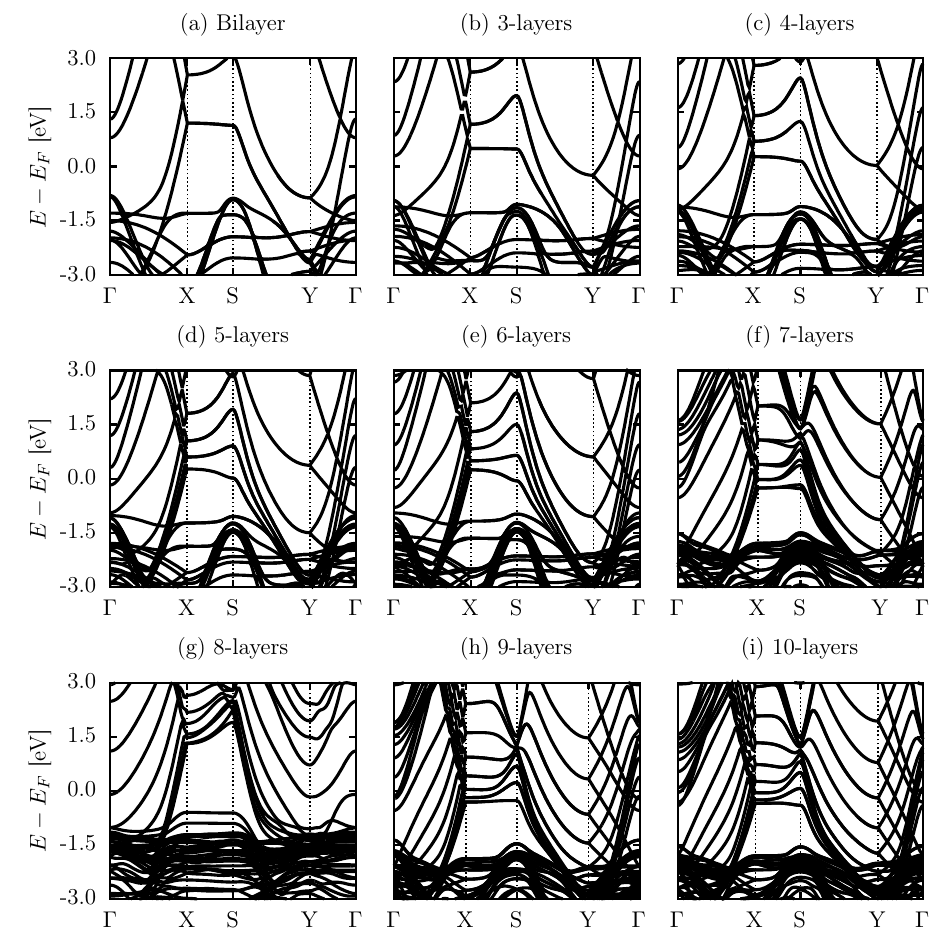}
    \caption{Electronic band structures of multilayered Goldene systems, ranging from bilayer up to 10 layers.}
    \label{fig:bands02}
\end{figure*}

Furthermore, the anisotropy of electronic transport in Goldene is a noteworthy feature. The band structure retains its metallic character along the $\Gamma$-X and Y-$\Gamma$ directions, promoting robust electronic conductivity. However, along the X-Y path, the band structure shows characteristics indicative of semiconducting behavior, suggesting that Goldene exhibits anisotropic electronic transport. This anisotropy could be exploited in designing devices with desired directional conductivity, offering new possibilities for tailoring Goldene’s electronic properties in practical applications.

\begin{figure*}[!htb]
    \centering
    \includegraphics[width=16cm]{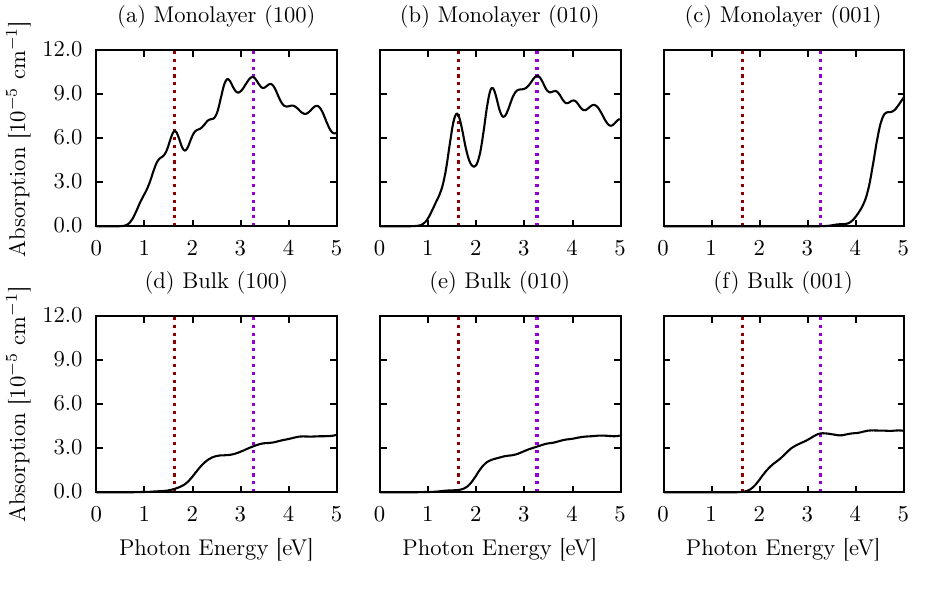}
    \caption{Optical absorption activity of (a-c) monolayer Goldene and (d-f) bulk-like Goldene for incident light polarized along the x- (100), y- (010), and z-directions (001). The dashed red and violet lines mark the visible range of the optical spectrum.}
    \label{fig:optical}
\end{figure*}

Figures \ref{fig:bands02}(a-i) present the electronic band structures of multilayered Goldene systems, ranging from bilayer up to 10 layers. As discussed in the previous section, the vacuum distance between the layers was fixed to avoid convergence toward bulk gold’s band structure, allowing for a more accurate exploration of the electronic behavior across a varying number of layers.

In examining the band structures of these multilayered systems, we observe that the primary electronic characteristics in the monolayer case remain primarily intact across all the multilayers. This consistency in band configuration highlights that the weak vdW interactions among the layers do not significantly modify the band structure imposed by the quantum confinement effects of the monolayer. As a result, the Dirac cones characteristic of monolayer Goldene are replicated across the stacked layers, with slight shifts due to interlayer coupling, as illustrated in Figures \ref{fig:bands02}(a-i). These shifts lead to superpositioned Dirac cones, a hallmark of layered 2D materials under weak vdW interactions. As the number of layers increases, the electronic states become more aligned with those of bulk-like gold (see Figure \ref{fig:bands01})(b), explaining the transition in the Dirac points.

It is worth mentioning that the multilayered systems exhibit three distinct groups in their energy dispersions. The band structures for the bilayer and the systems consisting of 3 to 6 layers maintain similar configurations (see Figures \ref{fig:bands02}(a-e)). In contrast, the 7- and 8-layer systems form another distinct grouping, as shown in Figures \ref{fig:bands02}(f-g)), and the 9- and 10-layer systems share a third configuration (refer to Figures \ref{fig:bands02}(h-i))). These variations can be traced back to the differences in lattice arrangements between these groups, which influence the interlayer coupling and, consequently, the electronic band structures. In particular, the stronger interlayer interactions in the thicker systems lead to more pronounced changes in the band structure. However, the overall metallic nature of Goldene is preserved across all cases.

The persistence of the metallic behavior across different numbers of layers, the slight shifts in Dirac cones, and the emergence of distinct electronic configurations in thicker layers suggest that Goldene's electronic properties can be finely tuned by varying the number of layers. This tunability, combined with the weak interlayer coupling, makes multilayered Goldene a promising candidate for applications that require adjustable electronic behavior while retaining the essential conductive properties of the monolayer.

Figures \ref{fig:optical}(a-c) and \ref{fig:optical}(d-f) display the optical absorption spectra for monolayer and bulk-like Goldene, respectively,  for the case of incident light polarized along the x- (100), y- (010), and z-directions (001). The dashed red and violet lines in these figures mark the visible range of the optical spectrum. These spectra provide crucial insights into different configurations of Goldene's anisotropic and isotropic optical behaviors. In the case of monolayer Goldene (Figures \ref{fig:optical}(a-c)), we observe intense absorption activity in the UV-Vis range for incident light polarized along the basal plane directions (x- and y-polarizations, see Figures \ref{fig:optical}(a-b)). The absorption coefficients for these directions reach values as high as 10$^{5}$ cm$^{-1}$, reflecting strong interaction with in-plane light. In contrast, the absorption in the out-of-plane (z-direction) polarization is limited to the UV range, with negligible absorption observed in the visible spectrum, as shown in Figure \ref{fig:optical}(c).

This distinct behavior stems from the anisotropic nature of the monolayer Goldene. The intense optical activity in the in-plane directions can be attributed to the highly delocalized electrons within the Goldene plane, which interact strongly with the incident light. This planar electronic configuration leads to robust dipole transitions when light is polarized along the basal plane, resulting in strong UV-Vis absorption. On the other hand, when the light is polarized along the z-direction (perpendicular to the plane), the absorption is significantly suppressed, with activity confined only to the UV-range. This suppression can be understood by the limited electronic states available for out-of-plane transitions in the monolayer. The absence of significant electronic overlap between layers in the z-direction decreases the interaction with light polarized along this axis, leading to much lower absorption in the UV-Vis range.

In contrast, bulk-like Goldene (Figures \ref{fig:optical}(d-f)) exhibits nearly isotropic optical absorption across all polarization directions. The absorption coefficients are lower in magnitude compared to the monolayer case, reflecting the influence of interlayer coupling in the bulk-like structure. Unlike the monolayer, where strong in-plane electron interactions dominate the optical response, the bulk-like phase introduces weak VDW interactions among layers, diluting the optical absorption.

This isotropic absorption in bulk-like Goldene can be attributed to the increased thickness and layer stacking, which allow for more uniform electronic interactions across all three directions (x-, y-, and z-polarizations). Consequently, the absorption in the x- and y-directions is decreased relative to the monolayer, as the interlayer coupling facilitates additional dipole transitions perpendicular to the basal plane. Consequently, the overall absorption magnitude remains smaller than in the monolayer, as the confined electronic states of the monolayer provide a more pronounced response to incident light.

The difference in optical absorption intensity between the monolayer and bulk-like Goldene is primarily due to the systems' dimensionality. The monolayer's reduced dimensionality results in more substantial quantum confinement effects, which enhance its interaction with incident light in the basal plane. In contrast, with its additional layers, the bulk-like phase experiences weakened confinement, leading to more moderate absorption intensities. This difference highlights the tunability of optical properties in Goldene, making it a promising candidate for applications in optoelectronics, where control over absorption characteristics is crucial.

\begin{figure}[!htb]
    \centering
    \includegraphics[width=8cm]{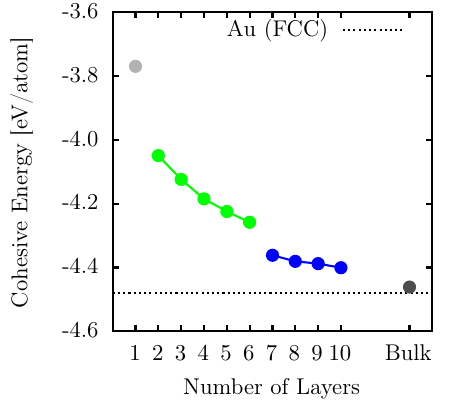}
    \caption{Multilayered Goldene's structural evolution as a layer number function, focusing on the cohesive energy. The dashed line refers to the cohesive energy for the FCC bulk gold.}
    \label{fig:celayers}
\end{figure}

Finally, we will discuss the structural properties of multilayered Goldene systems to answer the question: How does Goldene Stack? In this way, Figure \ref{fig:celayers} presents a comprehensive analysis of multilayered Goldene's structural evolution as a function of the number of layers, highlighting the cohesive energy and stacking configurations. This figure reveals how the stacking behavior of Goldene transitions through distinct phases, providing crucial insights into its structural stability and phase transitions. 

\begin{figure*}[!htb]
    \centering
    \includegraphics[width=16cm]{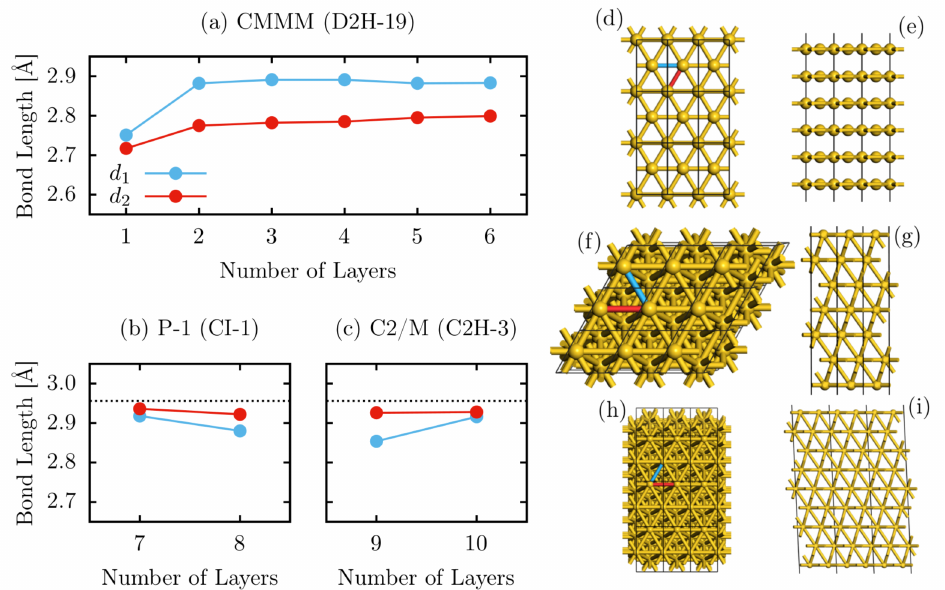}
    \caption{(a) Evolution of two characteristic bond lengths, denoted as $d_1$ (blue) and $d_2$ (red), for Goldene systems from one to six layers. Panels (b,c) show the bond lengths $d_1$ and $d_2$ for seven and eight layers and nine and ten, respectively. Panels (d,e), (f,g), and (h,i) show the top and side views of representative cases for each group of the lattice arrangement presented in Figure \ref{fig:celayers}, i.e., the 6-layers, the 7-layers, and the 9-layers cases, respectively.}
    \label{fig:stacking}
\end{figure*}

Goldene exhibits a consistent AA-like stacking configuration from bilayer to six layers, the green points in Figure \ref{fig:celayers}. This stacking arrangement, discussed in Figure \ref{fig:stacking} is characterized by a cohesive energy ranging between -4.26 and -4.09 eV/atom. Goldene maintains this AA-like arrangement throughout this range, unlike graphene, which adopts Bernal (AB) stacking in its multilayer forms \cite{wang2019ultrastiff,wang2016stacking}. This stability of AA-like stacking suggests that Goldene’s interlayer interactions are distinct from those in graphene, resulting in a different stacking behavior that is less prone to the formation of Bernal stacking. 

The AA-like stacking in Goldene implies a uniform alignment of the gold atoms across layers, leading to consistent electronic and structural properties of up to six layers. This behavior contrasts with other 2D materials, where different stacking configurations can significantly alter physical properties \cite{guo2021stacking,zhang2016van}. The observed stability of the AA-like stacking in Goldene highlights its unique structural characteristics. It influences its electronic and optical properties.

At seven layers, a notable structural change occurs where Goldene begins to transition to an FCC structure, accompanied by the emergence of rhombohedral stacking, as discussed later. This transition is reflected in the cohesive energy data, with a shift in the energy values and the onset of blue lines in Figure \ref{fig:celayers}, indicating a change in the stacking configuration. The cohesive energy for this phase converges to a range between -4.40 and -4.36 eV/atom, approaching that of bulk gold. The appearance of FCC-like stacking at this stage signifies a structural transition towards a more stable, bulk-like arrangement. The transition to FCC stacking aligns with the observed behavior of bulk gold, which is known for its FCC lattice structure. This shift underscores the material’s tendency to adopt stacking configurations that minimize the overall energy as the number of layers increases, reflecting a more stable bulk phase.

Beyond ten layers (Bulk-like Goldene), the lattice arrangement of Goldene fully converges to that of bulk gold. This convergence is represented by the consistent black circle in Figure \ref{fig:celayers} and signifies that the stacking behavior in these thicker multilayer configurations closely mirrors the FCC structure of bulk gold. The cohesive energy stabilizes within the range of around -4.46 eV/atom, reinforcing the resemblance to bulk gold’s structural properties (-4.48 eV/atom). The convergence to a bulk-like FCC structure indicates that Goldene’s multilayer stacking eventually reaches a configuration that is energetically favorable and consistent with the known properties of bulk gold. This transition highlights the material’s adaptability and ability to maintain structural integrity as the number of layers increases.

Analyzing the stacking behavior in multilayered Goldene provides critical insights into its structural evolution and stability. The persistent AA-like stacking of up to six layers suggests a unique interlayer interaction specific to Goldene, differentiating it from other 2D materials like graphene. The transition to FCC stacking at seven layers and the eventual convergence to bulk gold’s structure reflects the material’s tendency to adopt configurations that align with lower-energy, bulk-like phases as the number of layers increases. These findings have significant implications for understanding Goldene’s properties and potential applications. The distinctive stacking behavior and its transition to a bulk-like FCC structure offer valuable information for designing and utilizing Goldene in various technological applications. Additionally, the stability of different stacking configurations provides insights into the material’s mechanical and electronic properties, which are crucial for practical applications in electronics and materials science.

Figure \ref{fig:stacking}(a) shows the evolution of two characteristic bond lengths, denoted as $d_1$ (blue) and $d_2$ (red), for Goldene systems from one to six layers. Initially, the bond lengths for the monolayer are $d_1=2.75$ \r{A} $d_2 = 2.72$ \r{A}, which slightly increase for the bilayer configuration, reaching $d_1=2.88$ \r{A} $d_2 = 2.77$ \r{A}. This bond length evolution continues as the number of layers increases, converging at $d_1=2.88$ \r{A} $d_2 = 2.80$ \r{A} for systems with three up to six layers. This convergence in bond lengths reflects the structural stability of multilayered Goldene up to six layers. The changes in bond lengths are minor and signify that stacking additional layers does not drastically alter the bonding interactions between gold atoms, preserving the lattice symmetry.

For the lattice parameters, $l_{0x}$ and $l_{0y}$, the monolayer values are 2.89 \r{A} and 4.69 \r{A}, respectively. The bilayer shows similar trends, with lattice parameters of $l_{0x}=2.82$ \r{A} and $l_{0y}=4.78$ \r{A}. As the system progresses to three to six layers, these values converge to $l_{0x}=2.88$ \r{A} and $l_{0y}=4.79$ \r{A}, with the lattice angles relaxing to $\alpha=\beta=\gamma=90^{\circ}$. These values are in excellent agreement with the reported experimental data for monolayer and bilayer Goldene \cite{kashiwaya2024synthesis}, as well as other theoretical calculations for monolayer Goldene, which report similar lattice parameters of $l_{0x}=2.82$ \r{A} and $l_{0y}=4.78$ \r{A} \cite{zhao2024electrical,mortazavi2024goldene}. Up to six layers of the space group for the monolayer and multilayer Goldene crystal structure are identified as CMMM (D2H-19). This space group is commonly found in layered 2D materials, where the symmetry reflects the anisotropy in the in-plane and out-of-plane directions. For example, another 2D material with CMMM symmetry is multilayered phosphorene \cite{ribeiro2015group}.

Figures \ref{fig:stacking}(b) and \ref{fig:stacking}(c) show the bond lengths $d_1$ and $d_2$ for seven and eight layers and for nine and ten layers, respectively. The general trend observed in these figures is that the bond lengths approach the values typical of FCC bulk gold, represented by the black dashed line. However, minor fluctuations are observed in both the seven-layer and nine-layer systems without significant changes in structural symmetry due to numerical error propagation.

A distinct symmetry group emerges for the seven- and eight-layer systems: P-1 (CI-1). This symmetry is characteristic of rhombohedral stacking, which marks the transition from AA-like stacking to a more bulk-like arrangement. The rhombohedral structure reflects a lower symmetry compared to the simple AA-like stacking, as layers are no longer perfectly aligned but instead exhibit a slight offset. This behavior is consistent with the structural transitions observed in other layered materials like rhombohedral graphite \cite{nery2020long,xu2012pathway}. The symmetry shifts to C2/M (C2H-3) for the nine- and ten-layer systems. This symmetry group indicates a monoclinic structure, where the layers are stacked in an alternating pattern similar to the ABC stacking of bulk gold. This transition to C2/M symmetry indicates that Goldene converges fully to a bulk-like arrangement of FFC gold crystal. Examples of other 2D materials exhibiting C2/M symmetry include certain transition metal dichalcogenides in their bulk forms \cite{molina2015vibrational,guo2020prediction}. Importantly, the interlayer distance is slightly reduced to 2.79 \r{A} for the systems with three to six layers. In contrast, for the bulk-like ABC-stacked systems (nine and ten layers), the interlayer distance decreases further to 2.41–2.46 \r{A}. This decrease in the interlayer spacing is typical of 2D materials as they transition from few-layered structures to bulk-like phases, where interlayer interactions (such as VDW forces) become more relevant.

Figures \ref{fig:stacking}(d) and \ref{fig:stacking}(e) illustrate the top and side views of the six-layered Goldene system. The uniform AA stacking is evident in these panels, where the gold atoms in each layer are perfectly aligned, resulting in a high degree of symmetry and uniformity across the structure. The persistence of AA-like stacking up to six layers confirms the stability of this configuration in Goldene, which contrasts with the Bernal (AB) stacking found in multilayer graphene, as mentioned above. Figures \ref{fig:stacking}(f) and  \ref{fig:stacking}(g) depict the seven-layered Goldene system. Here, a transition from AA-like stacking to ABC-like stacking is observed, characterized by rhombohedral stacking. This transition is marked by a slight offset between the layers, breaking the AA-stacked system's symmetry and introducing a more complex lattice arrangement. The rhombohedral stacking pattern is consistent with the symmetry group P-1 (CI-1), reflecting the beginning of a transition towards a bulk-like structure. Finally, Figures \ref{fig:stacking}(h) and  \ref{fig:stacking}(i) show the nine-layered Goldene system, where the stacking arrangement has fully transitioned to the ABC stacking characteristic of bulk gold. In this configuration, the lattice arrangement is more closely aligned with the FCC structure, and the interlayer distances are consistent with those of bulk gold. This transition marks the final convergence of Goldene to a bulk-like phase, with the structure adopting the stable and energetically favorable ABC stacking pattern.

\section{Conclusions}

In summary, we carried out a comprehensive study on multilayered Goldene's structural and energetic properties, a recently synthesized monolayer of gold with atomic thickness. Our analysis has uncovered Goldene's unique stacking behavior and stability as the number of layers increases, offering valuable insights into its transition from a few-layered system to a bulk-like phase.

Goldene's monolayer and few-layered forms demonstrate exceptional stability, with cohesive energy and bond lengths following consistent and predictable trends. Goldene maintains an AA-like stacking pattern of up to six layers, a unique feature that sets it apart from the Bernal (AB) stacking typically observed in multilayer graphene. This distinctive stacking behavior underscores the unique structural properties of Goldene compared to other 2D materials. 

A significant transition occurs when the system reaches seven layers, at which point Goldene adopts a rhombohedral stacking configuration. This change marks the onset of a bulk-like phase and is characterized by a shift in symmetry, yielding lattices with distinct space groups. This stacking transition, similar to that observed in other layered materials, such as rhombohedral graphite, highlights the structural adaptability of Goldene as the number of layers increases. A progressive transition to the ABC stacking configuration typical of FCC bulk gold was also observed beyond seven layers. When the system reaches nine and ten layers, its structural arrangement, interlayer distances, and bond lengths converge to those of bulk gold.

In addition to the structural and energetic properties, we thoroughly investigated Goldene's electronic and optical characteristics. The monolayer Goldene exhibited a metallic nature with a single Dirac cone at the X-point of the Brillouin zone, a key feature preserved across multilayered systems, albeit with energy shifts as the number of layers increased. The multilayered cases present two Dirac cones in the band structure at the X- and Y-points. The metallic behavior observed in monolayer and bulk-like Goldene is marked by a high Fermi velocity, indicating its potential for efficient electronic transport. Furthermore, the electronic band structure displayed anisotropic pathways for electron conduction, particularly along specific high-symmetry directions in the Brillouin zone.

The optical properties of Goldene further underline its unique behavior, with significant absorption activity in the UV-Vis range for in-plane (x,y) light polarization. In contrast, the out-of-plane (z) polarization showed activity only in the UV range. This anisotropic optical absorption behavior is a defining characteristic of the monolayer. At the same time, the bulk-like Goldene exhibits more isotropic but considerably weaker optical absorption. The strong absorption in the visible range for the monolayer points to its potential applications in optoelectronic devices, making it an attractive material for further exploration in light-harvesting technologies. 

\begin{acknowledgments}
This work received partial support from Brazilian agencies CAPES, CNPq, and FAPDF.
M.L.P.J. acknowledges the financial support of the FAP-DF grant 00193-00001807/2023-16. The authors also acknowledge support from CENAPAD-SP (National High-Performance Center in São Paulo, State University of Campinas -- UNICAMP, project: proj950 and proj960) and NACAD (High-Performance Computing Center, Lobo Carneiro Supercomputer, Federal University of Rio de Janeiro -- UFRJ, project: a22002) for the computational support provided. 

L.A.R.J. acknowledges the financial support from FAP-DF grants 00193.00001808/2022-71 and 00193-00001857/2023-95, FAPDF-PRONEM grant 00193.00001247/2021-20, and CNPq grants 350176/2022-1 and 167745/2023-9. L.A.R.J also acknowledges PDPG-FAPDF-CAPES Centro-Oeste grant number $00193-00000867/2024-94$.
D. S. G. acknowledges the Center for Computing in Engineering and Sciences at Unicamp for financial support through the FAPESP/CEPID Grant \#2013/08293-7.
\end{acknowledgments}

\appendix

% The \nocite command causes all entries in a bibliography to be printed out
% whether or not they are actually referenced in the text. This is appropriate
% for the sample file to show the different styles of references, but authors
% most likely will not want to use it.
\nocite{*}

\bibliography{bibliography.bib}
\end{document}